%% file: FPCP04-hep.tex
\documentclass[12pt]{article}
\usepackage{epsfig,graphicx, eepic, epic, cite}
\usepackage{fpcp04}

\newcommand{\beq}{\begin{equation}}
\newcommand{\eeq}{\end{equation}}
\def\bea{\begin{eqnarray}}
\def\eea{\end{eqnarray}}
\def\nn{\nonumber}
\def\bec{\begin{center}}
\def\eec{\end{center}}

\input fpcp04-macro
\begin{document}

\Title{ Possibility of Large EW Penguin contribution
        \footnote{ Talk presented by T. Yoshikawa at International
        Conference on Flaver Physics and CP violation (FPCP2004),
        Daegu Korea, Oct. 4-9,
        2004 and YITP workshop on CP Violation and Matter and
        Anti-Matter Asymmetry (YITP-W-04-21), Kyoto, Japan,
        Jan. 12-14, 2005. 
        These
        talks were based on the work \cite{Y2}. }}
\bigskip

%%%%%%%%%%%%%%%%%%%%%%%%%%%%%%%%%%%%%
% Label to flag the first page of your contribution
% Replace Perret by your name starting with a capital letter
%
\label{TYoshikawa}

%%%%%%%%%%%%%%%%%%%%%%%%%%%%%%%%%%%%%
% Your name
%
\author{ Satoshi Mishima$~{}^a$ ~{\rm and} 
        ~~Tadashi Yoshikawa$~{}^b$ \index{T. Yoshikawa} }

%%%%%%%%%%%%%%%%%%%%%%%%%%%%%%%%%%%%%
% Your address
%
\address{a) Department of Physics, 
Tohoku University Sendai 980-8578, Japan \\ 
~b) Department of Physics, 
Nagoya University Nagoya 464-8602, Japan 
}

\makeauthor
\abstracts{
We discuss  
a possibility of large electroweak (EW) penguin contribution in 
$B\rightarrow K \pi$ and $\pi \pi$.  
The recent experimental data may be still suggesting that there are some
discrepancies between the data and theoretical estimations. 
In $B\rightarrow K\pi$ decays, to explain several 
theoretical relations among the branching ratios, 
a slightly large electroweak penguin contribution and large
strong phase differences or quite large color suppressed tree
contribution seem to be needed.    
The contributions should appear also in $B\rightarrow \pi\pi$.   
We show, as an example, a solution to solve 
the discrepancies in both $B\rightarrow K\pi$ and $B\rightarrow
\pi\pi$. It may be suggesting to need the large electroweak penguin
contribution with new weak phases and some SU(3) breaking effects by
new physics in both QCD and electroweak penguin type processes.     
}
$B\rightarrow K\pi $ modes have already been
measured well~(See 
the web page by Heavy Flavor Averaging Group~\cite{HFAG}) and they will 
be useful informations to understand the CP violation through 
the Kobayashi-Maskawa (KM) phases. 
If we can directly solve them by using the branching ratios and CP
asymmetries, it is very elegant way to 
determine the parameters and the weak phases. 
However, before to do so, 
it seemed to be slightly difficult to explain several 
relations among the branching ratios of $B\rightarrow K\pi$
and  $B\rightarrow \pi\pi$ without a large
electroweak (EW) penguin contribution with large
phase\cite{Lip,Y1,Y2,GROROS,BFRS,CL,WZ,HM,BL}. 
After ICHEP04 conference,
the data was slightly updated and the difficulty was quite
relaxed. However the situation we need some larger contributions than
the theoretical estimations did not change and furthermore a
discrepancy among the direct CP asymmetries remained.    
To satisfy these data, we find that 
the role of a color-favored EW penguin or color-suppressed tree 
may be important. The role of the EW penguin was pointed out 
and their magnitude was
estimated in several works~\cite{BF,NEU,GNK,GPY,FM}.   
They said that the ratio between 
gluonic and EW penguins is about $0.14$,
but the experimental data may suggest that the magnitude seems to be slightly 
larger than the estimation~\cite{Y2,GROROS,BFRS,CL}. 
Furthermore, one of the most difficult points to
explain them is that we will need quite large strong phase difference of EW
penguin diagram compared with the other strong phases. It is difficult
to produce the such large strong phase in the SM. 
If there is quite large deviation in the contribution 
from the EW penguin, it may suggest a possibility of new physics in 
these modes. In the usual sense, 
new physics contributions should be through in some
loop effects such as the penguin-type diagram so that there should not be
any discrepancies in tree-type diagrams. We put the new physics
contributions with weak phase into both gluonic and EW penguins to 
find the allowed regions for each parameters.

Using the diagram decomposition
approach~\cite{GHLR1,FMB,BF,NEU,FM} and redefinition of
the parameters,
the decay amplitudes of $B \rightarrow K^x\pi^y$ and $\pi^x \pi^y $,
$A_K^{xy}$ and $A_\pi^{xy}$, are written as follows: 
\bea
A^{0+}_K &=& - P |V_{tb}^*V_{ts}|
         \left[ 1 - r_A e^{i\delta^A}e^{i\phi_3}  
                                        \right]  , \\
A^{00}_K &=& 
     - \frac{1}{\sqrt{2}} P |V_{tb}^*V_{ts}|
         \left[ 1 - r_{EW}e^{i\delta^{EW}}  
             + r_C e^{i\delta^C}e^{i\phi_3}
                                        \right]  , \\
A^{+-}_K &=&
            P |V_{tb}^*V_{ts}| 
            \left[ 1 + r_{EW}^C  e^{i\delta^{EWC}} 
             - r_T e^{i\delta^T}e^{i\phi_3}
                                        \right]  , \\
A^{+0}_K &=&  \frac{1}{\sqrt{2}}
            P |V_{tb}^*V_{ts}|
                      \left[ 1 + r_{EW} e^{i\delta^{EW}} 
                               + r_{EW}^C e^{i\delta^{EWC}} 
             - ( r_Te^{i\delta^T} + r_C e^{i\delta^C} + r_A e^{i\delta^A} )
                       e^{i\phi_3} 
                                        \right]  , 
\eea
\bea
A^{00}_\pi &=& 
     \frac{1}{\sqrt{2}} T |V_{ub}^*V_{ud}|
         \left[ (\tilde{r}_P e^{-i\delta^{T}} 
              - \tilde{r}_{EW} e^{i(\delta^{EW}-\delta^{T})})e^{-i\phi_1}  
             - (\tilde{r}_C e^{i(\delta^C-\delta^T)} 
                - \tilde{r}_E e^{i(\delta^E-\delta^T)} )e^{i\phi_3}
                                        \right], \\
A^{+-}_\pi &=&
            - T |V_{ub}^*V_{ud}| 
            \left[ (\tilde{r}_P e^{-i\delta^{T}} + \tilde{r}_{EW}^C
         e^{i(\delta^{EWC} - \delta^T) })e^{-i\phi_1}   
            + (1 + \tilde{r}_E e^{i(\delta^E-\delta^T)}) e^{i\phi_3}
                                        \right]  , \\
A^{+0}_\pi &=&  -\frac{1}{\sqrt{2}}
           T|V_{ub}^*V_{ud}|
                      \left[ (\tilde{r}_{EW} e^{i(\delta^{EW}-\delta^T)} 
                               + \tilde{r}_{EW}^C
         e^{i(\delta^{EWC}-\delta^T)}
                ) e^{-i\phi_1}  
         + (1 + \tilde{r}_C e^{i(\delta^C-\delta^T)} )
                       e^{i\phi_3} 
                                        \right], 
\eea 
where $\phi_1$ and $\phi_3$ are the weak phases, 
$\delta^X$'s are the strong phase differences between 
each diagram and gluonic penguin, and 
$ r_T = \frac{ |T V_{ub}^*V_{us}| }{ |P V_{tb}^*V_{ts}|}\;, 
r_C = \frac{ |C V_{ub}^*V_{us}| }{ |P V_{tb}^*V_{ts}|}\;,
r_A = \frac{ |A V_{ub}^*V_{us}| }{ |P V_{tb}^*V_{ts}|}\;, 
r_{EW} = \frac{ |P_{EW}|}{ |P|}\;, 
r_{EW}^C = \frac{ |P_{EW}^C|}{ |P|}\;,
\tilde{r}_P = \frac{|P V_{tb}^*V_{td}|}{|T V_{ub}^*V_{ud}|} =
\frac{1}{r_T} \frac{|V_{td}V_{us}|}{|V_{ud}V_{ts}|}\;, 
\tilde{r}_C = \frac{|C|}{|T|} = \frac{r_C}{r_T}\;,  
\tilde{r}_E = \frac{|E|}{|T|}\;,  
\tilde{r}_{EW} = \frac{|P_{EW} V_{tb}^*V_{td}|}{|T V_{ub}^*V_{ud}|}
                = r_{EW} \tilde{r}_P\;,
\tilde{r}_{EW}^C = \frac{|P_{EW}^C V_{tb}^*V_{td}|}{|T
V_{ub}^*V_{ud}|} = r_{EW}^C \tilde{r}_P\;, $
where $T$ is a color-favored tree amplitude, $C$ is a color-suppressed 
tree, $A$($E$) is an annihilation (exchange), $P$ 
is a gluonic penguin, $P_{EW}$ is a
color-favored EW penguin and $P_{EW}^C$ is a color-suppressed 
EW penguin. To discuss the dependence of each diagram, 
we assume the hierarchy of the ratios as
$ 1 > r_T, r_{EW} > r_C, r_{EW}^C > r_A $ and 
$ 1 > \tilde{r}_P > \tilde{r}_{EW}, 
\tilde{r}_C > \tilde{r}_{EW}^C, \tilde{r}_E $~\cite{GHLR1}. 
$r_T$ can be estimated as $r_T \sim 0.2 $ with $10\%$ error 
from the the ratio of $Br(B^+ \rightarrow \pi^0
\pi^+)$ to $Br(B^+ \rightarrow K^0
\pi^+)$.  
$r_C$ and $r_{EW}^C$ must be suppressed by color factor from $r_T$ 
and $r_{EW}$ and 
we can assume that  $r_C \sim 0.1\, r_{T}$ and $r_{EW}^C \sim 0.1\,
r_{EW}$. $r_A$ could be negligible because 
$B$ meson decay constant works as a suppression factor $f_B/M_B$. 
While, by the similar way one can obtain $\tilde{r}_P \sim 0.3$,
$\tilde{r}_C = 0.1$.
Indeed,
the estimations for each parameters in the PQCD approach\cite{KLS,US} are $
r_T =  0.21,~ r_{EW} = 0.14, ~ 
r_C =  0.018, ~ r^C_{EW} = 0.012 $ and $
r_A = 0.0048. $ 
According to this assumption, 
we neglect the $r^2$ terms including $r_C, r_A$ and $r_{EW}^C$ in 
$B\rightarrow K\pi $. 
In $B\rightarrow \pi\pi $ we will neglect $\tilde{r}_{EW}^C$ 
and $\tilde{r}_E$ for simplicity, but keep $\tilde{r}_{EW}$ to discuss
its magnitude in the both modes.        

Under the assumptions, one can find several relations among the
branching ratios for $B\rightarrow K\pi$ decays up to $O(r)$. 
Here we list the 3 relations as follows:  
\bea
R_c-R_n &= & \frac{2 \bar{B}_K^{+0}}{\bar{B}_K^{0+}} 
    - \frac{\bar{B}_K^{+-}}{2\bar{B}_K^{00}} \nn \\ 
        &=& 
    - 2 r_{EW}^2 \cos2\delta^{EW} 
    + 2 r_{EW} r_{T} \cos(\delta^{EW}+\delta^{T})\cos\phi_3
    \, =\, 0.21 \pm 0.11\;,
\label{B+0B0+MB+-B00} \\[2mm]
S ~~~~  &= & \frac{2 \bar{B}_K^{+0}}{\bar{B}_K^{0+}}
 - \frac{\tau^+}{\tau^0}\frac{\bar{B}_K^{+-}}{\bar{B}_K^{0+}} + 
\frac{\tau^+}{\tau^0}\frac{2 \bar{B}_K^{00}}{\bar{B}_K^{0+}} -1 \nn \\
   &=& 
2 r_{EW}^2 - 2 r_{EW} r_{T} \cos(\delta^{EW}-\delta^{T})\cos\phi_3\, 
                                                      =\, 0.22 \pm 0.14\;, 
\label{B+0MB+-PB00M1} \\[2mm]
R_+ -2 ~ &=& 
\frac{\tau^0}{\tau^+}\frac{2\bar{B}_K^{+0}}{\bar{B}_K^{+-}} + 
\frac{\tau^+}{\tau^0}\frac{2\bar{B}_K^{00}}{\bar{B}_K^{0+}} - 2 \nn \\
&=& 2 r_{EW}^2 
          + 2 r_{EW} r_{T} \cos(\delta^{EW}+\delta^{T})\cos\phi_3\, 
                                                      =\, 0.26 \pm 0.16\;, 
\label{B+0MB00M2}
\eea 
where $\bar{B}_K^{xy}$ shows the branching ratio for $B\rightarrow K^x
\pi^y $ decay.  If we can neglect all $r^2$ terms for the theoretical 
estimations $r \sim  O(0.1)$, these relations 
should be close to zero 
but the experimental data do not seem to satisfy them.  
Thus it may show there is existing a discrepancy between theoretical
estimations and experimental data. The difference comes from $r^2$
term including $r_{EW}$ 
so that one can find these deviations may be an evidence that the EW
penguin is larger than the estimation we expected within the SM. 

What we can expect at present are roughly $40^\circ < \phi_3 <
80^\circ $ from CKM fitting and $r_T = 0.2$ from 
the ratio $\frac{B_\pi^{+0}}{B_K^{0+}}$.
Hence from the old experimental data $r_{EW}$ would be larger than 
0.2 while the theoretical prediction of $r_{EW}$ is 0.14, and a large 
strong phase difference between gluonic and EW penguins will be 
requested due to keep the positive $R_c-R_n$~\cite{Y2}. 
Accordingly, to explain the data we may need some contribution from new
physics in the EW-penguin-type contribution with a large phase.  
Under exact flavor SU(3) symmetry, the strong phase difference between
the EW penguin and the color-favored tree, which is called as $\omega$,
$(\omega \equiv \delta^{EW}-\delta^T )$, 
should be close to zero because the diagrams are topologically same~\cite{NEU}
and effectively the difference is whether just only the exchanging weak
gauge boson is $W$ or $Z$. 
If it is correct, the constraint for
$\delta^T$ has to influence on $\delta^{EW}$ due to $\delta^{EW} \sim
\delta^T$.   
We consider the direct CP asymmetry to obtain the informations
about strong phase.       

The direct CP asymmetries of $B\rightarrow K\pi $ 
under the same assumption which we neglect the
terms of $O(0.001)$ are
\bea
A_{CP}^{0+} &\equiv & \frac{|A_K^{0-}|^2 - |A_K^{0+}|^2}{|A_K^{0-}|^2 
                + |A_K^{0+}|^2}\,
             =\, - 2 r_A \sin\delta^A \sin\phi_3\,= -0.020\pm0.034 , \\[0.5mm]
A_{CP}^{00} &\equiv & \frac{|\bar{A}_K^{00}|^2 - |A_K^{00}|^2}
                     {|\bar{A}_K^{00}|^2 + |A_K^{00}|^2}\,
             =\, 2 r_C \sin\delta^C \sin\phi_3\, = -0.09\pm0.14 \;, \\[0.5mm]
A_{CP}^{+-} &\equiv & \frac{|A_K^{-+}|^2 
                      - |A_K^{+-}|^2}{|A_K^{-+}|^2 + |A_K^{+-}|^2}\,
             =\, - 2 r_T \sin\delta^T \sin\phi_3
             - r_T^2 \sin2\delta^T \sin2\phi_3  = -0.109 \pm 0.019
                              \\[0.5mm]
A_{CP}^{+0} &\equiv & \frac{|A_K^{-0}|^2 - |A_K^{+0}|^2}
                           {|A_K^{-0}|^2 + |A_K^{+0}|^2}\,
             =\, - 2 (r_T \sin\delta^T + r_C \sin\delta^C
                  + r_A \sin\delta^A )\sin\phi_3 \nn \\
            & & ~~~~~~~~~~~~
         + 2 r_{EW} r_{T} \sin(\delta^T + \delta^{EW} )\sin\phi_3
               - r_T^2 \sin2\delta^T \sin2\phi_3\; = 0.04 \pm 0.04. 
\eea
Up to the order of $r^2$, there is a
relation among the CP asymmetries as follows:
\bea  
A_{CP}^{+0} - A_{CP}^{+-} + A_{CP}^{00} - A_{CP}^{0+}\, =\, 
         2 r_T  r_{EW} \sin(\delta^T + \delta^{EW}) \sin\phi_3\, 
                                               =\, 0.08\pm 0.15\;. 
\eea
The discrepancy of this relation from zero 
is caused by the cross term of $r_T$
and $r_{EW}$. This may also give us some useful informations about 
$r_{EW}$ and the strong phases but the data of $A_{CP}^{00}$ still has 
quite large error, so that one can not extract from it
at present time. If we can neglect $r_C$ and $r_A$ according to the 
our hierarchy assumption, the relation will be 
$A_{CP}^{+0} - A_{CP}^{+-} = 
         2 r_T  r_{EW} \sin(\delta^T + \delta^{EW}) \sin\phi_3 
                                               = 0.15\pm 0.04 $ so
that it seems to say also $r_{EW}$ should be larger value than the
usual estimation. The difference between $A_{CP}^{+-}$ and
$A_{CP}^{+0}$ is also an important information to understand whether the
origin of the deviations is $r_{EW}$ or $r_C$.      

\begin{figure}[thb]
\bec
\hspace*{3mm}
\begin{minipage}[l]{2.9in} 
\begin{center}
% GNUPLOT: LaTeX picture using EEPIC macros
\setlength{\unitlength}{0.085450pt}
\begin{picture}(2400,1800)(0,0)
\footnotesize
\thicklines \path(370,249)(411,249)
\thicklines \path(2276,249)(2235,249)
\put(329,249){\makebox(0,0)[r]{ 0}}
\thicklines \path(370,494)(411,494)
\thicklines \path(2276,494)(2235,494)
\put(329,494){\makebox(0,0)[r]{ 0.1}}
\thicklines \path(370,739)(411,739)
\thicklines \path(2276,739)(2235,739)
\put(329,739){\makebox(0,0)[r]{ 0.2}}
\thicklines \path(370,984)(411,984)
\thicklines \path(2276,984)(2235,984)
\put(329,984){\makebox(0,0)[r]{ 0.3}}
\thicklines \path(370,1228)(411,1228)
\thicklines \path(2276,1228)(2235,1228)
\put(329,1228){\makebox(0,0)[r]{ 0.4}}
\thicklines \path(370,1473)(411,1473)
\thicklines \path(2276,1473)(2235,1473)
\put(329,1473){\makebox(0,0)[r]{ 0.5}}
\thicklines \path(370,1718)(411,1718)
\thicklines \path(2276,1718)(2235,1718)
\put(329,1718){\makebox(0,0)[r]{ 0.6}}
\thicklines \path(370,249)(370,290)
\thicklines \path(370,1718)(370,1677)
\put(370,166){\makebox(0,0){ 0.1}}
\thicklines \path(688,249)(688,290)
\thicklines \path(688,1718)(688,1677)
\put(688,166){\makebox(0,0){ 0.15}}
\thicklines \path(1005,249)(1005,290)
\thicklines \path(1005,1718)(1005,1677)
\put(1005,166){\makebox(0,0){ 0.2}}
\thicklines \path(1323,249)(1323,290)
\thicklines \path(1323,1718)(1323,1677)
\put(1323,166){\makebox(0,0){ 0.25}}
\thicklines \path(1641,249)(1641,290)
\thicklines \path(1641,1718)(1641,1677)
\put(1641,166){\makebox(0,0){ 0.3}}
\thicklines \path(1958,249)(1958,290)
\thicklines \path(1958,1718)(1958,1677)
\put(1958,166){\makebox(0,0){ 0.35}}
\thicklines \path(2276,249)(2276,290)
\thicklines \path(2276,1718)(2276,1677)
\put(2276,166){\makebox(0,0){ 0.4}}
\thicklines \path(370,249)(2276,249)(2276,1718)(370,1718)(370,249)
\thicklines \path(370,249)(2276,249)(2276,1718)(370,1718)(370,249)
\put(-210,1090){\makebox(0,0)[l]{{$R_c-R_n $}}}
\put(640,1250){\makebox(0,0)[l]{$2 \sigma $  }}
\put(640,790){\makebox(0,0)[l]{$1 \sigma $  }}
\put(1323,-25){\makebox(0,0){{\large $ r_{EW} $} }}
\put(1500,980){\makebox(0,0)[l]{$\phi_3 = 40^\circ $  }}
\put(1910,968){\makebox(0,0)[l]{$~~~~~60^\circ $  }}
\put(1910,860){\makebox(0,0)[l]{$~~~~~80^\circ $  }}
\thinlines \path(370,337)(370,337)(434,350)(497,364)(561,379)(624,396)(688,413)(751,431)(815,450)(878,470)(942,491)(1005,513)(1069,536)(1132,559)(1196,584)(1259,610)(1323,637)(1387,665)(1450,694)(1514,723)(1577,754)(1641,786)(1704,818)(1768,852)(1831,887)(1895,922)(1958,959)(2022,997)(2085,1035)(2149,1075)(2212,1115)(2276,1157)
\thinlines \path(370,317)(370,317)(434,329)(497,342)(561,356)(624,370)(688,386)(751,403)(815,420)(878,439)(942,458)(1005,479)(1069,500)(1132,523)(1196,546)(1259,571)(1323,596)(1387,623)(1450,650)(1514,678)(1577,708)(1641,738)(1704,769)(1768,802)(1831,835)(1895,869)(1958,904)(2022,941)(2085,978)(2149,1016)(2212,1055)(2276,1095)
\thinlines \path(370,304)(370,304)(434,314)(497,326)(561,339)(624,353)(688,367)(751,383)(815,400)(878,417)(942,436)(1005,455)(1069,476)(1132,498)(1196,520)(1259,544)(1323,568)(1387,594)(1450,620)(1514,647)(1577,676)(1641,705)(1704,736)(1768,767)(1831,799)(1895,833)(1958,867)(2022,902)(2085,938)(2149,976)(2212,1014)(2276,1053)
\thinlines \dashline[-20]{16}(370,1032)(370,1032)(389,1032)(409,1032)(428,1032)(447,1032)(466,1032)(486,1032)(505,1032)(524,1032)(543,1032)(563,1032)(582,1032)(601,1032)(620,1032)(640,1032)(659,1032)(678,1032)(697,1032)(717,1032)(736,1032)(755,1032)(774,1032)(794,1032)(813,1032)(832,1032)(851,1032)(871,1032)(890,1032)(909,1032)(928,1032)(948,1032)(967,1032)(986,1032)(1005,1032)(1025,1032)(1044,1032)(1063,1032)(1082,1032)(1102,1032)(1121,1032)(1140,1032)(1159,1032)(1179,1032)(1198,1032)(1217,1032)(1236,1032)(1256,1032)(1275,1032)(1294,1032)(1313,1032)
\thinlines \dashline[-20]{16}(1313,1032)(1333,1032)(1352,1032)(1371,1032)(1390,1032)(1410,1032)(1429,1032)(1448,1032)(1467,1032)(1487,1032)(1506,1032)(1525,1032)(1544,1032)(1564,1032)(1583,1032)(1602,1032)(1621,1032)(1641,1032)(1660,1032)(1679,1032)(1698,1032)(1718,1032)(1737,1032)(1756,1032)(1775,1032)(1795,1032)(1814,1032)(1833,1032)(1852,1032)(1872,1032)(1891,1032)(1910,1032)(1929,1032)(1949,1032)(1968,1032)(1987,1032)(2006,1032)(2026,1032)(2045,1032)(2064,1032)(2083,1032)(2103,1032)(2122,1032)(2141,1032)(2160,1032)(2180,1032)(2199,1032)(2218,1032)(2237,1032)(2257,1032)(2276,1032)
\thinlines \dashline[-20]{16}(370,494)(370,494)(389,494)(409,494)(428,494)(447,494)(466,494)(486,494)(505,494)(524,494)(543,494)(563,494)(582,494)(601,494)(620,494)(640,494)(659,494)(678,494)(697,494)(717,494)(736,494)(755,494)(774,494)(794,494)(813,494)(832,494)(851,494)(871,494)(890,494)(909,494)(928,494)(948,494)(967,494)(986,494)(1005,494)(1025,494)(1044,494)(1063,494)(1082,494)(1102,494)(1121,494)(1140,494)(1159,494)(1179,494)(1198,494)(1217,494)(1236,494)(1256,494)(1275,494)(1294,494)(1313,494)
\thinlines \dashline[-20]{16}(1313,494)(1333,494)(1352,494)(1371,494)(1390,494)(1410,494)(1429,494)(1448,494)(1467,494)(1487,494)(1506,494)(1525,494)(1544,494)(1564,494)(1583,494)(1602,494)(1621,494)(1641,494)(1660,494)(1679,494)(1698,494)(1718,494)(1737,494)(1756,494)(1775,494)(1795,494)(1814,494)(1833,494)(1852,494)(1872,494)(1891,494)(1910,494)(1929,494)(1949,494)(1968,494)(1987,494)(2006,494)(2026,494)(2045,494)(2064,494)(2083,494)(2103,494)(2122,494)(2141,494)(2160,494)(2180,494)(2199,494)(2218,494)(2237,494)(2257,494)(2276,494)
\thinlines \dashline[-20]{11}(370,1302)(370,1302)(389,1302)(409,1302)(428,1302)(447,1302)(466,1302)(486,1302)(505,1302)(524,1302)(543,1302)(563,1302)(582,1302)(601,1302)(620,1302)(640,1302)(659,1302)(678,1302)(697,1302)(717,1302)(736,1302)(755,1302)(774,1302)(794,1302)(813,1302)(832,1302)(851,1302)(871,1302)(890,1302)(909,1302)(928,1302)(948,1302)(967,1302)(986,1302)(1005,1302)(1025,1302)(1044,1302)(1063,1302)(1082,1302)(1102,1302)(1121,1302)(1140,1302)(1159,1302)(1179,1302)(1198,1302)(1217,1302)(1236,1302)(1256,1302)(1275,1302)(1294,1302)(1313,1302)
\thinlines \dashline[-20]{11}(1313,1302)(1333,1302)(1352,1302)(1371,1302)(1390,1302)(1410,1302)(1429,1302)(1448,1302)(1467,1302)(1487,1302)(1506,1302)(1525,1302)(1544,1302)(1564,1302)(1583,1302)(1602,1302)(1621,1302)(1641,1302)(1660,1302)(1679,1302)(1698,1302)(1718,1302)(1737,1302)(1756,1302)(1775,1302)(1795,1302)(1814,1302)(1833,1302)(1852,1302)(1872,1302)(1891,1302)(1910,1302)(1929,1302)(1949,1302)(1968,1302)(1987,1302)(2006,1302)(2026,1302)(2045,1302)(2064,1302)(2083,1302)(2103,1302)(2122,1302)(2141,1302)(2160,1302)(2180,1302)(2199,1302)(2218,1302)(2237,1302)(2257,1302)(2276,1302)
\end{picture}
\end{center} 
\end{minipage} 
\hspace*{5mm}
\begin{minipage}[r]{2.9in}
\begin{center}
% GNUPLOT: LaTeX picture using EEPIC macros
\setlength{\unitlength}{0.085450pt}
\begin{picture}(2400,1800)(0,0)
\footnotesize
\thicklines \path(370,249)(411,249)
\thicklines \path(2276,249)(2235,249)
\put(329,249){\makebox(0,0)[r]{ 0}}
\thicklines \path(370,494)(411,494)
\thicklines \path(2276,494)(2235,494)
\put(329,494){\makebox(0,0)[r]{ 0.1}}
\thicklines \path(370,739)(411,739)
\thicklines \path(2276,739)(2235,739)
\put(329,739){\makebox(0,0)[r]{ 0.2}}
\thicklines \path(370,984)(411,984)
\thicklines \path(2276,984)(2235,984)
\put(329,984){\makebox(0,0)[r]{ 0.3}}
\thicklines \path(370,1228)(411,1228)
\thicklines \path(2276,1228)(2235,1228)
\put(329,1228){\makebox(0,0)[r]{ 0.4}}
\thicklines \path(370,1473)(411,1473)
\thicklines \path(2276,1473)(2235,1473)
\put(329,1473){\makebox(0,0)[r]{ 0.5}}
\thicklines \path(370,1718)(411,1718)
\thicklines \path(2276,1718)(2235,1718)
\put(329,1718){\makebox(0,0)[r]{ 0.6}}
\thicklines \path(370,249)(370,290)
\thicklines \path(370,1718)(370,1677)
\put(370,166){\makebox(0,0){ 0}}
\thicklines \path(635,249)(635,290)
\thicklines \path(635,1718)(635,1677)
\put(635,166){\makebox(0,0){ 50}}
\thicklines \path(899,249)(899,290)
\thicklines \path(899,1718)(899,1677)
\put(899,166){\makebox(0,0){ 100}}
\thicklines \path(1164,249)(1164,290)
\thicklines \path(1164,1718)(1164,1677)
\put(1164,166){\makebox(0,0){ 150}}
\thicklines \path(1429,249)(1429,290)
\thicklines \path(1429,1718)(1429,1677)
\put(1429,166){\makebox(0,0){ 200}}
\thicklines \path(1694,249)(1694,290)
\thicklines \path(1694,1718)(1694,1677)
\put(1694,166){\makebox(0,0){ 250}}
\thicklines \path(1958,249)(1958,290)
\thicklines \path(1958,1718)(1958,1677)
\put(1958,166){\makebox(0,0){ 300}}
\thicklines \path(2223,249)(2223,290)
\thicklines \path(2223,1718)(2223,1677)
\put(2223,166){\makebox(0,0){ 350}}
\thicklines \path(370,249)(2276,249)(2276,1718)(370,1718)(370,249)
\put(-230,850){\makebox(0,0)[l]{{ $R_c-R_n$}  }}
\put(433,770){\makebox(0,0)[l]{$1 \sigma $  }}
\put(433,450){\makebox(0,0)[l]{$2 \sigma $  }}
\put(433,1228){\makebox(0,0)[l]{$2 \sigma $  }}
\put(1480,1000){\makebox(0,0)[l]{$r_{EW} = 0.40 $  }}
\put(1620,690){\makebox(0,0)[l]{$~~~~~0.30 $  }}
\put(1610,530){\makebox(0,0)[l]{$~~~~~0.20 $  }}
\put(1610,300){\makebox(0,0)[l]{$~~~~~0.14 $  }}
\put(1283,42){\makebox(0,0){{$ \delta^{EW} $} }}
\thinlines \path(370,254)(370,254)(375,254)(381,253)(386,253)(391,253)(396,253)(402,253)(407,253)(412,253)(418,253)(423,253)(428,254)(434,254)(439,255)(444,255)(449,256)(455,257)(460,258)(465,258)(471,259)(476,260)(481,262)(486,263)(492,264)(497,265)(502,267)(508,268)(513,269)(518,271)(524,272)(529,274)(534,276)(539,277)(545,279)(550,281)(555,282)(561,284)(566,286)(571,288)(576,289)(582,291)(587,293)(592,295)(598,296)(603,298)(608,300)(614,302)(619,303)(624,305)(629,307)
\thinlines \path(629,307)(635,308)(640,310)(645,312)(651,313)(656,315)(661,316)(666,317)(672,319)(677,320)(682,321)(688,323)(693,324)(698,325)(704,326)(709,327)(714,328)(719,329)(725,330)(730,331)(735,331)(741,332)(746,332)(751,333)(756,334)(762,334)(767,335)(772,335)(778,335)(783,336)(788,337)(794,338)(799,339)(804,339)(809,340)(815,340)(820,340)(825,340)(831,340)(836,340)(841,340)(847,340)(852,339)(857,339)(862,338)(868,337)(873,336)(878,335)(884,334)(889,333)(894,331)
\thinlines \path(894,331)(899,330)(905,328)(910,327)(915,325)(921,323)(926,321)(931,319)(937,317)(942,314)(947,312)(952,310)(958,307)(963,304)(968,302)(974,299)(979,296)(984,293)(989,290)(995,287)(1000,284)(1005,281)(1011,278)(1016,274)(1021,271)(1027,268)(1032,264)(1037,261)(1042,257)(1048,254)(1053,250)(1058,249)(1064,249)(1069,249)(1074,249)(1079,249)(1085,249)(1090,249)(1095,249)(1101,249)(1106,249)(1111,249)(1117,249)(1122,249)(1127,249)(1132,249)(1138,249)(1143,249)(1148,249)(1154,249)(1159,249)
\thinlines \path(1159,249)(1164,249)(1169,249)(1175,249)(1180,249)(1185,249)(1191,249)(1196,249)(1201,249)(1207,249)(1212,249)(1217,249)(1222,249)(1228,249)(1233,249)(1238,249)(1244,249)(1249,249)(1254,249)(1259,249)(1265,249)(1270,249)(1275,249)(1281,249)(1286,249)(1291,249)(1297,249)(1302,249)(1307,249)(1312,249)(1318,249)(1323,249)(1328,249)(1334,249)(1339,249)(1344,249)(1349,249)(1355,249)(1360,249)(1365,249)(1371,249)(1376,249)(1381,249)(1387,249)(1392,249)(1397,249)(1402,249)(1408,249)(1413,249)(1418,249)(1424,249)
\thinlines \path(1424,249)(1429,249)(1434,249)(1439,249)(1445,249)(1450,249)(1455,249)(1461,249)(1466,249)(1471,249)(1477,249)(1482,249)(1487,249)(1492,249)(1498,249)(1503,249)(1508,249)(1514,249)(1519,249)(1524,249)(1529,249)(1535,249)(1540,249)(1545,249)(1551,249)(1556,249)(1561,249)(1567,249)(1572,249)(1577,249)(1582,253)(1588,256)(1593,260)(1598,264)(1604,267)(1609,271)(1614,274)(1619,278)(1625,281)(1630,285)(1635,288)(1641,292)(1646,295)(1651,298)(1657,301)(1662,304)(1667,307)(1672,310)(1678,314)(1683,318)(1688,322)
\thinlines \path(1688,322)(1694,326)(1699,330)(1704,334)(1709,338)(1715,341)(1720,345)(1725,348)(1731,352)(1736,355)(1741,358)(1747,361)(1752,364)(1757,367)(1762,369)(1768,372)(1773,374)(1778,376)(1784,379)(1789,381)(1794,383)(1800,384)(1805,386)(1810,387)(1815,389)(1821,390)(1826,391)(1831,392)(1837,393)(1842,394)(1847,394)(1852,395)(1858,395)(1863,395)(1868,396)(1874,395)(1879,395)(1884,395)(1890,395)(1895,394)(1900,394)(1905,393)(1911,392)(1916,391)(1921,390)(1927,389)(1932,388)(1937,386)(1942,385)(1948,383)(1953,381)
\thinlines \path(1953,381)(1958,380)(1964,378)(1969,376)(1974,374)(1980,372)(1985,370)(1990,368)(1995,365)(2001,363)(2006,361)(2011,358)(2017,356)(2022,353)(2027,351)(2032,348)(2038,346)(2043,343)(2048,340)(2054,338)(2059,335)(2064,332)(2070,329)(2075,327)(2080,324)(2085,321)(2091,319)(2096,316)(2101,313)(2107,310)(2112,308)(2117,305)(2122,302)(2128,300)(2133,297)(2138,295)(2144,292)(2149,290)(2154,287)(2160,285)(2165,283)(2170,281)(2175,278)(2181,276)(2186,274)(2191,273)(2197,271)(2202,269)(2207,268)(2212,266)(2218,265)
\thinlines \path(2218,265)(2223,263)(2228,262)(2234,261)(2239,260)(2244,259)(2250,258)(2255,257)(2260,256)(2265,255)(2271,255)(2276,254)
\thinlines \path(370,249)(370,249)(375,249)(381,249)(386,249)(391,249)(396,249)(402,249)(407,249)(412,249)(418,249)(423,249)(428,249)(434,249)(439,249)(444,249)(449,249)(455,249)(460,249)(465,249)(471,249)(476,249)(481,249)(486,249)(492,249)(497,249)(502,249)(508,249)(513,249)(518,249)(524,251)(529,255)(534,259)(539,264)(545,268)(550,272)(555,276)(561,281)(566,285)(571,290)(576,294)(582,299)(587,303)(592,308)(598,313)(603,317)(608,322)(614,326)(619,331)(624,335)(629,340)
\thinlines \path(629,340)(635,344)(640,348)(645,353)(651,357)(656,361)(661,365)(666,369)(672,373)(677,376)(682,380)(688,383)(693,387)(698,390)(704,393)(709,396)(714,399)(719,402)(725,405)(730,408)(735,410)(741,412)(746,415)(751,417)(756,419)(762,421)(767,422)(772,424)(778,425)(783,427)(788,429)(794,431)(799,433)(804,434)(809,435)(815,436)(820,437)(825,438)(831,438)(836,438)(841,438)(847,437)(852,437)(857,436)(862,434)(868,433)(873,431)(878,430)(884,427)(889,425)(894,423)
\thinlines \path(894,423)(899,420)(905,417)(910,414)(915,410)(921,406)(926,403)(931,399)(937,394)(942,390)(947,385)(952,381)(958,376)(963,370)(968,365)(974,360)(979,354)(984,348)(989,342)(995,336)(1000,330)(1005,324)(1011,317)(1016,311)(1021,304)(1027,297)(1032,291)(1037,284)(1042,277)(1048,270)(1053,263)(1058,256)(1064,249)(1069,249)(1074,249)(1079,249)(1085,249)(1090,249)(1095,249)(1101,249)(1106,249)(1111,249)(1117,249)(1122,249)(1127,249)(1132,249)(1138,249)(1143,249)(1148,249)(1154,249)(1159,249)
\thinlines \path(1159,249)(1164,249)(1169,249)(1175,249)(1180,249)(1185,249)(1191,249)(1196,249)(1201,249)(1207,249)(1212,249)(1217,249)(1222,249)(1228,249)(1233,249)(1238,249)(1244,249)(1249,249)(1254,249)(1259,249)(1265,249)(1270,249)(1275,249)(1281,249)(1286,249)(1291,249)(1297,249)(1302,249)(1307,249)(1312,249)(1318,249)(1323,249)(1328,249)(1334,249)(1339,249)(1344,249)(1349,249)(1355,249)(1360,249)(1365,249)(1371,249)(1376,249)(1381,249)(1387,249)(1392,249)(1397,249)(1402,249)(1408,249)(1413,249)(1418,249)(1424,249)
\thinlines \path(1424,249)(1429,249)(1434,249)(1439,249)(1445,249)(1450,249)(1455,249)(1461,249)(1466,249)(1471,249)(1477,249)(1482,249)(1487,249)(1492,249)(1498,249)(1503,249)(1508,249)(1514,249)(1519,249)(1524,249)(1529,249)(1535,249)(1540,249)(1545,249)(1551,249)(1556,249)(1561,249)(1567,249)(1572,249)(1577,255)(1582,262)(1588,270)(1593,277)(1598,284)(1604,291)(1609,298)(1614,305)(1619,312)(1625,319)(1630,326)(1635,333)(1641,339)(1646,346)(1651,352)(1657,358)(1662,364)(1667,370)(1672,376)(1678,382)(1683,390)(1688,397)
\thinlines \path(1688,397)(1694,404)(1699,411)(1704,418)(1709,424)(1715,431)(1720,437)(1725,443)(1731,449)(1736,454)(1741,459)(1747,464)(1752,469)(1757,474)(1762,478)(1768,482)(1773,486)(1778,489)(1784,493)(1789,496)(1794,499)(1800,501)(1805,503)(1810,505)(1815,507)(1821,509)(1826,510)(1831,511)(1837,512)(1842,512)(1847,513)(1852,513)(1858,512)(1863,512)(1868,511)(1874,510)(1879,509)(1884,507)(1890,506)(1895,504)(1900,502)(1905,499)(1911,497)(1916,494)(1921,491)(1927,488)(1932,485)(1937,481)(1942,477)(1948,473)(1953,469)
\thinlines \path(1953,469)(1958,465)(1964,461)(1969,456)(1974,452)(1980,447)(1985,442)(1990,437)(1995,432)(2001,426)(2006,421)(2011,415)(2017,410)(2022,404)(2027,399)(2032,393)(2038,387)(2043,381)(2048,375)(2054,370)(2059,364)(2064,358)(2070,352)(2075,346)(2080,340)(2085,334)(2091,328)(2096,322)(2101,317)(2107,311)(2112,305)(2117,300)(2122,294)(2128,289)(2133,283)(2138,278)(2144,273)(2149,268)(2154,263)(2160,258)(2165,253)(2170,249)(2175,249)(2181,249)(2186,249)(2191,249)(2197,249)(2202,249)(2207,249)(2212,249)(2218,249)
\thinlines \path(2218,249)(2223,249)(2228,249)(2234,249)(2239,249)(2244,249)(2250,249)(2255,249)(2260,249)(2265,249)(2271,249)(2276,249)
\thinlines \path(370,249)(370,249)(375,249)(381,249)(386,249)(391,249)(396,249)(402,249)(407,249)(412,249)(418,249)(423,249)(428,249)(434,249)(439,249)(444,249)(449,249)(455,249)(460,249)(465,249)(471,249)(476,249)(481,249)(486,249)(492,249)(497,249)(502,249)(508,249)(513,249)(518,249)(524,249)(529,249)(534,249)(539,249)(545,249)(550,249)(555,249)(561,251)(566,263)(571,275)(576,286)(582,298)(587,310)(592,322)(598,334)(603,346)(608,358)(614,370)(619,382)(624,394)(629,405)
\thinlines \path(629,405)(635,417)(640,429)(645,440)(651,451)(656,462)(661,473)(666,484)(672,494)(677,504)(682,514)(688,524)(693,534)(698,543)(704,552)(709,561)(714,569)(719,577)(725,585)(730,593)(735,600)(741,607)(746,613)(751,619)(756,626)(762,631)(767,636)(772,641)(778,645)(783,651)(788,656)(794,660)(799,665)(804,668)(809,671)(815,674)(820,676)(825,677)(831,678)(836,679)(841,679)(847,678)(852,677)(857,675)(862,673)(868,671)(873,667)(878,663)(884,659)(889,654)(894,649)
\thinlines \path(894,649)(899,643)(905,637)(910,630)(915,623)(921,615)(926,607)(931,598)(937,589)(942,579)(947,569)(952,559)(958,548)(963,537)(968,525)(974,513)(979,501)(984,488)(989,475)(995,462)(1000,448)(1005,435)(1011,420)(1016,406)(1021,392)(1027,377)(1032,362)(1037,347)(1042,331)(1048,316)(1053,300)(1058,285)(1064,269)(1069,253)(1074,249)(1079,249)(1085,249)(1090,249)(1095,249)(1101,249)(1106,249)(1111,249)(1117,249)(1122,249)(1127,249)(1132,249)(1138,249)(1143,249)(1148,249)(1154,249)(1159,249)
\thinlines \path(1159,249)(1164,249)(1169,249)(1175,249)(1180,249)(1185,249)(1191,249)(1196,249)(1201,249)(1207,249)(1212,249)(1217,249)(1222,249)(1228,249)(1233,249)(1238,249)(1244,249)(1249,249)(1254,249)(1259,249)(1265,249)(1270,249)(1275,249)(1281,249)(1286,249)(1291,249)(1297,249)(1302,249)(1307,249)(1312,249)(1318,249)(1323,249)(1328,249)(1334,249)(1339,249)(1344,249)(1349,249)(1355,249)(1360,249)(1365,249)(1371,249)(1376,249)(1381,249)(1387,249)(1392,249)(1397,249)(1402,249)(1408,249)(1413,249)(1418,249)(1424,249)
\thinlines \path(1424,249)(1429,249)(1434,249)(1439,249)(1445,249)(1450,249)(1455,249)(1461,249)(1466,249)(1471,249)(1477,249)(1482,249)(1487,249)(1492,249)(1498,249)(1503,249)(1508,249)(1514,249)(1519,249)(1524,249)(1529,249)(1535,249)(1540,249)(1545,249)(1551,249)(1556,249)(1561,249)(1567,249)(1572,257)(1577,274)(1582,290)(1588,306)(1593,322)(1598,337)(1604,353)(1609,369)(1614,384)(1619,399)(1625,414)(1630,429)(1635,444)(1641,458)(1646,472)(1651,486)(1657,499)(1662,512)(1667,525)(1672,538)(1678,551)(1683,566)(1688,580)
\thinlines \path(1688,580)(1694,594)(1699,608)(1704,621)(1709,634)(1715,646)(1720,658)(1725,669)(1731,680)(1736,691)(1741,701)(1747,710)(1752,719)(1757,727)(1762,735)(1768,742)(1773,749)(1778,755)(1784,761)(1789,766)(1794,770)(1800,774)(1805,777)(1810,780)(1815,782)(1821,784)(1826,785)(1831,786)(1837,786)(1842,785)(1847,784)(1852,782)(1858,780)(1863,777)(1868,774)(1874,770)(1879,766)(1884,761)(1890,756)(1895,750)(1900,744)(1905,737)(1911,730)(1916,722)(1921,714)(1927,706)(1932,697)(1937,688)(1942,678)(1948,668)(1953,657)
\thinlines \path(1953,657)(1958,647)(1964,636)(1969,624)(1974,613)(1980,601)(1985,588)(1990,576)(1995,563)(2001,551)(2006,537)(2011,524)(2017,511)(2022,497)(2027,484)(2032,470)(2038,456)(2043,442)(2048,428)(2054,414)(2059,400)(2064,387)(2070,373)(2075,359)(2080,345)(2085,331)(2091,318)(2096,304)(2101,291)(2107,277)(2112,264)(2117,252)(2122,249)(2128,249)(2133,249)(2138,249)(2144,249)(2149,249)(2154,249)(2160,249)(2165,249)(2170,249)(2175,249)(2181,249)(2186,249)(2191,249)(2197,249)(2202,249)(2207,249)(2212,249)(2218,249)
\thinlines \path(2218,249)(2223,249)(2228,249)(2234,249)(2239,249)(2244,249)(2250,249)(2255,249)(2260,249)(2265,249)(2271,249)(2276,249)
\thinlines \path(370,249)(370,249)(375,249)(381,249)(386,249)(391,249)(396,249)(402,249)(407,249)(412,249)(418,249)(423,249)(428,249)(434,249)(439,249)(444,249)(449,249)(455,249)(460,249)(465,249)(471,249)(476,249)(481,249)(486,249)(492,249)(497,249)(502,249)(508,249)(513,249)(518,249)(524,249)(529,249)(534,249)(539,249)(545,249)(550,249)(555,249)(561,249)(566,249)(571,249)(576,258)(582,281)(587,303)(592,326)(598,349)(603,372)(608,394)(614,417)(619,440)(624,462)(629,485)
\thinlines \path(629,485)(635,507)(640,529)(645,551)(651,572)(656,594)(661,615)(666,635)(672,655)(677,675)(682,694)(688,714)(693,732)(698,750)(704,768)(709,785)(714,802)(719,818)(725,833)(730,848)(735,863)(741,876)(746,889)(751,901)(756,913)(762,925)(767,935)(772,944)(778,953)(783,964)(788,973)(794,982)(799,989)(804,996)(809,1002)(815,1007)(820,1011)(825,1014)(831,1016)(836,1018)(841,1018)(847,1017)(852,1016)(857,1013)(862,1009)(868,1005)(873,1000)(878,993)(884,986)(889,978)(894,969)
\thinlines \path(894,969)(899,959)(905,948)(910,936)(915,923)(921,910)(926,896)(931,880)(937,865)(942,848)(947,830)(952,812)(958,793)(963,774)(968,753)(974,732)(979,711)(984,689)(989,666)(995,642)(1000,619)(1005,594)(1011,570)(1016,544)(1021,519)(1027,493)(1032,466)(1037,440)(1042,413)(1048,386)(1053,358)(1058,331)(1064,303)(1069,275)(1074,249)(1079,249)(1085,249)(1090,249)(1095,249)(1101,249)(1106,249)(1111,249)(1117,249)(1122,249)(1127,249)(1132,249)(1138,249)(1143,249)(1148,249)(1154,249)(1159,249)
\thinlines \path(1159,249)(1164,249)(1169,249)(1175,249)(1180,249)(1185,249)(1191,249)(1196,249)(1201,249)(1207,249)(1212,249)(1217,249)(1222,249)(1228,249)(1233,249)(1238,249)(1244,249)(1249,249)(1254,249)(1259,249)(1265,249)(1270,249)(1275,249)(1281,249)(1286,249)(1291,249)(1297,249)(1302,249)(1307,249)(1312,249)(1318,249)(1323,249)(1328,249)(1334,249)(1339,249)(1344,249)(1349,249)(1355,249)(1360,249)(1365,249)(1371,249)(1376,249)(1381,249)(1387,249)(1392,249)(1397,249)(1402,249)(1408,249)(1413,249)(1418,249)(1424,249)
\thinlines \path(1424,249)(1429,249)(1434,249)(1439,249)(1445,249)(1450,249)(1455,249)(1461,249)(1466,249)(1471,249)(1477,249)(1482,249)(1487,249)(1492,249)(1498,249)(1503,249)(1508,249)(1514,249)(1519,249)(1524,249)(1529,249)(1535,249)(1540,249)(1545,249)(1551,249)(1556,249)(1561,249)(1567,249)(1572,274)(1577,302)(1582,330)(1588,358)(1593,386)(1598,414)(1604,442)(1609,469)(1614,496)(1619,523)(1625,549)(1630,575)(1635,600)(1641,625)(1646,650)(1651,674)(1657,698)(1662,721)(1667,743)(1672,765)(1678,788)(1683,813)(1688,836)
\thinlines \path(1688,836)(1694,860)(1699,882)(1704,904)(1709,925)(1715,945)(1720,964)(1725,983)(1731,1000)(1736,1017)(1741,1033)(1747,1048)(1752,1062)(1757,1075)(1762,1087)(1768,1098)(1773,1108)(1778,1118)(1784,1126)(1789,1133)(1794,1140)(1800,1145)(1805,1149)(1810,1153)(1815,1155)(1821,1156)(1826,1157)(1831,1156)(1837,1155)(1842,1152)(1847,1149)(1852,1144)(1858,1139)(1863,1133)(1868,1125)(1874,1117)(1879,1108)(1884,1098)(1890,1087)(1895,1076)(1900,1063)(1905,1050)(1911,1036)(1916,1021)(1921,1005)(1927,989)(1932,972)(1937,954)(1942,936)(1948,917)(1953,897)
\thinlines \path(1953,897)(1958,877)(1964,856)(1969,835)(1974,813)(1980,791)(1985,769)(1990,746)(1995,722)(2001,698)(2006,674)(2011,650)(2017,625)(2022,601)(2027,576)(2032,550)(2038,525)(2043,500)(2048,475)(2054,449)(2059,424)(2064,398)(2070,373)(2075,348)(2080,323)(2085,298)(2091,273)(2096,249)(2101,249)(2107,249)(2112,249)(2117,249)(2122,249)(2128,249)(2133,249)(2138,249)(2144,249)(2149,249)(2154,249)(2160,249)(2165,249)(2170,249)(2175,249)(2181,249)(2186,249)(2191,249)(2197,249)(2202,249)(2207,249)(2212,249)(2218,249)
\thinlines \path(2218,249)(2223,249)(2228,249)(2234,249)(2239,249)(2244,249)(2250,249)(2255,249)(2260,249)(2265,249)(2271,249)(2276,249)
\thinlines \dashline[-20]{16}(370,1032)(370,1032)(389,1032)(409,1032)(428,1032)(447,1032)(466,1032)(486,1032)(505,1032)(524,1032)(543,1032)(563,1032)(582,1032)(601,1032)(620,1032)(640,1032)(659,1032)(678,1032)(697,1032)(717,1032)(736,1032)(755,1032)(774,1032)(794,1032)(813,1032)(832,1032)(851,1032)(871,1032)(890,1032)(909,1032)(928,1032)(948,1032)(967,1032)(986,1032)(1005,1032)(1025,1032)(1044,1032)(1063,1032)(1082,1032)(1102,1032)(1121,1032)(1140,1032)(1159,1032)(1179,1032)(1198,1032)(1217,1032)(1236,1032)(1256,1032)(1275,1032)(1294,1032)(1313,1032)
\thinlines \dashline[-20]{16}(1313,1032)(1333,1032)(1352,1032)(1371,1032)(1390,1032)(1410,1032)(1429,1032)(1448,1032)(1467,1032)(1487,1032)(1506,1032)(1525,1032)(1544,1032)(1564,1032)(1583,1032)(1602,1032)(1621,1032)(1641,1032)(1660,1032)(1679,1032)(1698,1032)(1718,1032)(1737,1032)(1756,1032)(1775,1032)(1795,1032)(1814,1032)(1833,1032)(1852,1032)(1872,1032)(1891,1032)(1910,1032)(1929,1032)(1949,1032)(1968,1032)(1987,1032)(2006,1032)(2026,1032)(2045,1032)(2064,1032)(2083,1032)(2103,1032)(2122,1032)(2141,1032)(2160,1032)(2180,1032)(2199,1032)(2218,1032)(2237,1032)(2257,1032)(2276,1032)
\thinlines \dashline[-20]{16}(370,494)(370,494)(389,494)(409,494)(428,494)(447,494)(466,494)(486,494)(505,494)(524,494)(543,494)(563,494)(582,494)(601,494)(620,494)(640,494)(659,494)(678,494)(697,494)(717,494)(736,494)(755,494)(774,494)(794,494)(813,494)(832,494)(851,494)(871,494)(890,494)(909,494)(928,494)(948,494)(967,494)(986,494)(1005,494)(1025,494)(1044,494)(1063,494)(1082,494)(1102,494)(1121,494)(1140,494)(1159,494)(1179,494)(1198,494)(1217,494)(1236,494)(1256,494)(1275,494)(1294,494)(1313,494)
\thinlines \dashline[-20]{16}(1313,494)(1333,494)(1352,494)(1371,494)(1390,494)(1410,494)(1429,494)(1448,494)(1467,494)(1487,494)(1506,494)(1525,494)(1544,494)(1564,494)(1583,494)(1602,494)(1621,494)(1641,494)(1660,494)(1679,494)(1698,494)(1718,494)(1737,494)(1756,494)(1775,494)(1795,494)(1814,494)(1833,494)(1852,494)(1872,494)(1891,494)(1910,494)(1929,494)(1949,494)(1968,494)(1987,494)(2006,494)(2026,494)(2045,494)(2064,494)(2083,494)(2103,494)(2122,494)(2141,494)(2160,494)(2180,494)(2199,494)(2218,494)(2237,494)(2257,494)(2276,494)
\thinlines \dashline[-20]{11}(370,1302)(370,1302)(389,1302)(409,1302)(428,1302)(447,1302)(466,1302)(486,1302)(505,1302)(524,1302)(543,1302)(563,1302)(582,1302)(601,1302)(620,1302)(640,1302)(659,1302)(678,1302)(697,1302)(717,1302)(736,1302)(755,1302)(774,1302)(794,1302)(813,1302)(832,1302)(851,1302)(871,1302)(890,1302)(909,1302)(928,1302)(948,1302)(967,1302)(986,1302)(1005,1302)(1025,1302)(1044,1302)(1063,1302)(1082,1302)(1102,1302)(1121,1302)(1140,1302)(1159,1302)(1179,1302)(1198,1302)(1217,1302)(1236,1302)(1256,1302)(1275,1302)(1294,1302)(1313,1302)
\thinlines \dashline[-20]{11}(1313,1302)(1333,1302)(1352,1302)(1371,1302)(1390,1302)(1410,1302)(1429,1302)(1448,1302)(1467,1302)(1487,1302)(1506,1302)(1525,1302)(1544,1302)(1564,1302)(1583,1302)(1602,1302)(1621,1302)(1641,1302)(1660,1302)(1679,1302)(1698,1302)(1718,1302)(1737,1302)(1756,1302)(1775,1302)(1795,1302)(1814,1302)(1833,1302)(1852,1302)(1872,1302)(1891,1302)(1910,1302)(1929,1302)(1949,1302)(1968,1302)(1987,1302)(2006,1302)(2026,1302)(2045,1302)(2064,1302)(2083,1302)(2103,1302)(2122,1302)(2141,1302)(2160,1302)(2180,1302)(2199,1302)(2218,1302)(2237,1302)(2257,1302)(2276,1302)
\end{picture}
\eec
\end{minipage} 
\caption{The lines show the maximum bound of $R_c-R_n$ for $r_{EW}$ and
    $\delta^{EW} $
    at $r_T = 0.2 $ and $40^\circ < \phi_3 < 80^\circ $ 
    under constraint $-0.128 < A_{CP}^{+-} < -0.090$ }
    \label{fig:4}
\end{center}
\end{figure}
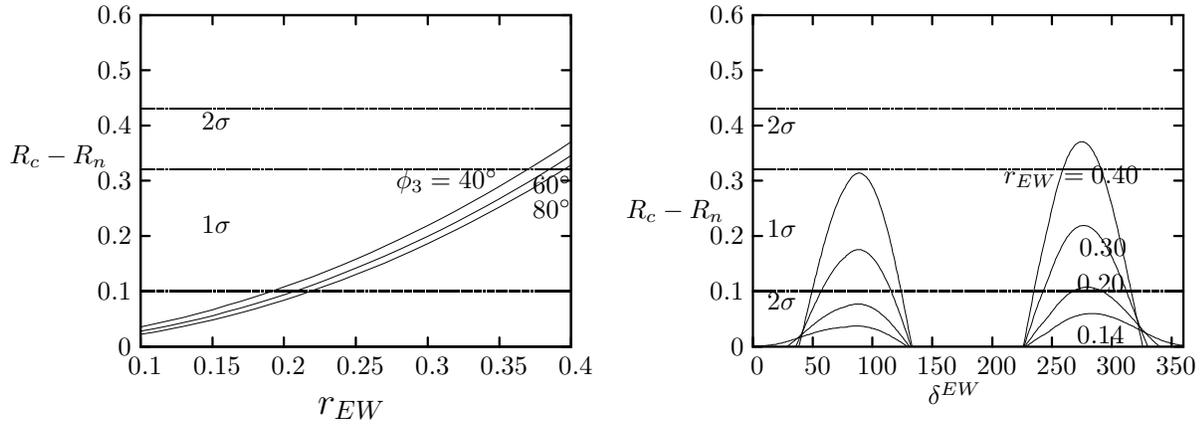

We use only $A_{CP}^{+-}$ because it is an
accurate measurement and will give a constraint to 
$\delta^T $.  Using both constraints from $A_{CP}^{+-} =
-0.109\pm0.019 $ and Fleischer-Mannel bound~\cite{FMB}, 
$R \equiv  \frac{\tau^+}{\tau^0}\frac{\bar{B}_K^{+-}}{\bar{B}_K^{0+}} 
= 1 - 2 r_{T} \cos\delta^{T} \cos\phi_3  
    + r_{T}^2 =\, 0.82\pm0.06~, $ one can find the
small $\delta^T$ is favored and $\delta^T $ should be around $15^\circ $. 
Taking account of these constraints for $\delta^T$ from $A_{CP}^{+-}$, we plot
the maximum bound of $R_c-R_n$ as the functions of  
$\delta^{EW}$ and $r_{EW}$ in Fig.~\ref{fig:4}, respectively. 
They show that at $1\sigma $ level $r_{EW}$ should be larger than
$0.2$ which is slightly larger than theoretical estimation $0.14$. 
Then the allowed regions for 
$\delta^{EW}-\delta^{T} $ around $0^\circ$ and $180^\circ$
disappear. $R_c-R_n$ seems to favor  
$45^\circ < |\delta^{EW}|<135^\circ$, but the constraint from
$A_{CP}^{+-}$ is strongly suggesting that the strong phase, $\delta^T$, 
should be around $15^\circ$. In consequence, $\delta^{EW}-\delta^{T} =
0$ as the theoretical prospect is disfavored.    
What the quite large strong phase difference is requested may be a
serious problem in these modes. 
If SU(3) symmetry is good one,
these properties should also appear in $B\rightarrow \pi \pi $.  

When we consider the ratios among the branching ratios 
for $B\rightarrow \pi \pi$ decays, 
\bea
\frac{2 \bar{B}_\pi^{00}}{\bar{B}_\pi^{+-}} = \frac{ 
   \tilde{r}_C^2 + \tilde{r}_P^2 (1 + r_{EW}^2 - 2 r_{EW}
   \cos\delta^{EW}) - 2 \tilde{r}_P \tilde{r}_C (\cos\delta^T  
        - {r}_{EW}\cos\omega ) \cos(\phi_1+\phi_3 ) }
   { 1 + \tilde{r}_P^2 + 2 \tilde{r}_P \cos\delta^T \cos(\phi_1+\phi_3
   ) }\;,  \\
\frac{\tau^0}{\tau^+}\frac{2
   \bar{B}_\pi^{+0}}{\bar{B}_\pi^{+-}} 
    = \frac{ 
   1 + \tilde{r}_C^2 + 2 \tilde{r}_C 
%\cos(\delta^C-\delta^T) 
     + \tilde{r}_P^2 {r}_{EW}^2  + 2 \tilde{r}_P {r}_{EW} (\cos\omega   
        + {r}_{C}\cos\omega ) \cos(\phi_1+\phi_3 ) }
   { 1 + \tilde{r}_P^2 + 2 \tilde{r}_P \cos\delta^T
     \cos(\phi_1+\phi_3)}\;,  
\eea   
there are also discrepancies between theoretical expectation and
experimental data. In above equations, $\delta^C$ is taken to be equal
to $\delta^T$. 
The theoretical rough estimations are  
$\frac{\tau^0}{\tau^+}\frac{2 \bar{B}_\pi^{0+}}{\bar{B}_\pi^{+-}} \sim
1 $ and $\frac{\bar{B}_\pi^{00}}{\bar{B}_\pi^{+-}} \sim
0.1 $ for $\tilde{r}_P \sim 0.3$, but the experimental data are quite large
values, $\frac{\tau^0}{\tau^+}\frac{2 \bar{B}_\pi^{0+}}{\bar{B}_\pi^{+-}} =
2.20\pm0.31 $ and $\frac{\bar{B}_\pi^{00}}{\bar{B}_\pi^{+-}} = 
0.66\pm0.13 $ and they are not consistent with the theoretical
expectations. To explain the discrepancy, 
the denominator seems to have to be smaller
value so that $\phi_1+\phi_3 $
should be larger than $90^\circ $. 
By the constraint from CKM fitting, $|\phi_1+\phi_3 |$ can not be so
much larger than $90^\circ $ that this may suggest some new phases is 
existing in these contributions.     
However it is not enough to explain the differences and 
we will also have to take account of SU(3) breaking effect. 
We can find that to explain 
the discrepancy, $b-d$ gluonic penguin contribution $P_\pi$
should be larger than $b-s$ gluonic penguin $P_K$ without the CKM
factor. It shows SU(3) breaking effect must appear in these decay
modes.  In addition, large EW penguin contribution also help to
enhance the ratios. Their ratios are enhanced by
$\tilde{r}_P$, $\tilde{r}_C$ and $r_{EW}$. However $\tilde{r}_C = C/T$ 
is $0.1$ for the naive estimation by factorization and it will be
at largest up to $1/N_c \sim 0.3$.
Large $\tilde{r}_P$ is an evidence to explain the discrepancies but it 
also has some constraints from $B\rightarrow KK$ decays which are pure
$b-d$ gluonic penguin processes. The constraint to $P_\pi/P_K $ comes from $
\frac{Br(B^0\rightarrow K^0 \bar{K}^0 )}
     {Br(B^+\rightarrow{K}^0 \pi^+)}\, \sim\, 
\frac{|P_\pi~V_{tb}^*V_{td}|^2 }{|P_K~V_{tb}^*V_{ts}|^2 }\, <\, 7.3 \times
10^{-2}\; $  so that $
\frac{P_\pi}{P_K}\, <\, 1.5\;. $ 
Thus $\tilde{r}_P$ may be allowed up to $0.3 \times 1.5 = 0.45
$.  

It is slightly difficult to get the values within the $1\sigma
$ region unless larger $r_C$ is allowed. However we feel that it may be
unnatural that such tree diagram obtains the larger contribution than
usual estimation. Therefore we consider the case keeping small $r_C$ and 
including some new effects in penguin contribution.

\begin{figure}[htb]
\bec
\hspace*{0mm}
\begin{minipage}[l]{2.75in} 
\begin{center}
% GNUPLOT: LaTeX picture using EEPIC macros
\setlength{\unitlength}{0.0820450pt}
\begin{picture}(2400,1800)(0,0)
\footnotesize
\thicklines \path(411,249)(452,249)
\thicklines \path(2276,249)(2235,249)
\put(370,249){\makebox(0,0)[r]{ 0}}
\thicklines \path(411,396)(452,396)
\thicklines \path(2276,396)(2235,396)
\put(370,396){\makebox(0,0)[r]{}}
\thicklines \path(411,543)(452,543)
\thicklines \path(2276,543)(2235,543)
\put(370,543){\makebox(0,0)[r]{ 0.1}}
\thicklines \path(411,690)(452,690)
\thicklines \path(2276,690)(2235,690)
\put(370,690){\makebox(0,0)[r]{}}
\thicklines \path(411,837)(452,837)
\thicklines \path(2276,837)(2235,837)
\put(370,837){\makebox(0,0)[r]{ 0.2}}
\thicklines \path(411,984)(452,984)
\thicklines \path(2276,984)(2235,984)
\put(370,984){\makebox(0,0)[r]{}}
\thicklines \path(411,1130)(452,1130)
\thicklines \path(2276,1130)(2235,1130)
\put(370,1130){\makebox(0,0)[r]{ 0.3}}
\thicklines \path(411,1277)(452,1277)
\thicklines \path(2276,1277)(2235,1277)
\put(370,1277){\makebox(0,0)[r]{}}
\thicklines \path(411,1424)(452,1424)
\thicklines \path(2276,1424)(2235,1424)
\put(370,1424){\makebox(0,0)[r]{ 0.4}}
\thicklines \path(411,1571)(452,1571)
\thicklines \path(2276,1571)(2235,1571)
\put(370,1571){\makebox(0,0)[r]{}}
\thicklines \path(411,1718)(452,1718)
\thicklines \path(2276,1718)(2235,1718)
\put(370,1718){\makebox(0,0)[r]{ 0.5}}
\thicklines \path(411,249)(411,290)
\thicklines \path(411,1718)(411,1677)
\put(411,166){\makebox(0,0){ 0}}
\thicklines \path(598,249)(598,290)
\thicklines \path(598,1718)(598,1677)
\put(598,166){\makebox(0,0){ }}
\thicklines \path(784,249)(784,290)
\thicklines \path(784,1718)(784,1677)
\put(784,166){\makebox(0,0){ 0.1}}
\thicklines \path(971,249)(971,290)
\thicklines \path(971,1718)(971,1677)
\put(971,166){\makebox(0,0){ }}
\thicklines \path(1157,249)(1157,290)
\thicklines \path(1157,1718)(1157,1677)
\put(1157,166){\makebox(0,0){ 0.2}}
\thicklines \path(1344,249)(1344,290)
\thicklines \path(1344,1718)(1344,1677)
\put(1344,166){\makebox(0,0){}}
\thicklines \path(1530,249)(1530,290)
\thicklines \path(1530,1718)(1530,1677)
\put(1530,166){\makebox(0,0){ 0.3}}
\thicklines \path(1717,249)(1717,290)
\thicklines \path(1717,1718)(1717,1677)
\put(1717,166){\makebox(0,0){}}
\thicklines \path(1903,249)(1903,290)
\thicklines \path(1903,1718)(1903,1677)
\put(1903,166){\makebox(0,0){ 0.4}}
\thicklines \path(2090,249)(2090,290)
\thicklines \path(2090,1718)(2090,1677)
\put(2090,166){\makebox(0,0){}}
\thicklines \path(2276,249)(2276,290)
\thicklines \path(2276,1718)(2276,1677)
\put(2276,166){\makebox(0,0){ 0.5}}
\thicklines \path(411,249)(2276,249)(2276,1718)(411,1718)(411,249)
\put(-82,983){\makebox(0,0)[l]{{\large $r_{C}$}}}
\put(1343,42){\makebox(0,0){\large $r_{EW}$ }}
\put(1443,1082){\makebox(0,0){\large Allowed region  }}
\put(1903,1792){\makebox(0,0){$\delta^{EW}, \delta^{C} $ are free.  }}
\thinlines \path(411,748)(411,748)(448,748)(486,748)(523,748)(560,719)(598,690)(635,690)(672,660)(709,631)(747,602)(784,572)(821,543)(859,513)(896,484)(933,455)(970,425)(1008,367)(1045,337)(1082,308)(1120,278)(1157,249)(1194,249)(1232,249)(1269,249)(1306,249)(1344,249)(1381,249)(1418,249)(1455,249)(1493,249)(1530,249)(1567,249)(1605,249)(1642,249)(1679,249)(1716,249)(1754,249)(1791,249)(1828,249)(1866,249)(1903,249)(1940,249)(1978,249)(2015,249)(2052,249)(2090,249)(2127,278)(2164,308)(2201,367)(2239,396)
\thinlines \path(2239,396)(2276,455)
\end{picture}
\end{center} 
\end{minipage} 
\hspace*{5mm}
\begin{minipage}[r]{2.75in}
\begin{center}
% GNUPLOT: LaTeX picture using EEPIC macros
\setlength{\unitlength}{0.0820450pt}
\begin{picture}(2400,1800)(0,0)
\footnotesize
\thicklines \path(411,249)(452,249)
\thicklines \path(2276,249)(2235,249)
\put(370,249){\makebox(0,0)[r]{ 0}}
\thicklines \path(411,396)(452,396)
\thicklines \path(2276,396)(2235,396)
\put(370,396){\makebox(0,0)[r]{}}
\thicklines \path(411,543)(452,543)
\thicklines \path(2276,543)(2235,543)
\put(370,543){\makebox(0,0)[r]{ 0.1}}
\thicklines \path(411,690)(452,690)
\thicklines \path(2276,690)(2235,690)
\put(370,690){\makebox(0,0)[r]{}}
\thicklines \path(411,837)(452,837)
\thicklines \path(2276,837)(2235,837)
\put(370,837){\makebox(0,0)[r]{ 0.2}}
\thicklines \path(411,984)(452,984)
\thicklines \path(2276,984)(2235,984)
\put(370,984){\makebox(0,0)[r]{}}
\thicklines \path(411,1130)(452,1130)
\thicklines \path(2276,1130)(2235,1130)
\put(370,1130){\makebox(0,0)[r]{ 0.3}}
\thicklines \path(411,1277)(452,1277)
\thicklines \path(2276,1277)(2235,1277)
\put(370,1277){\makebox(0,0)[r]{ }}
\thicklines \path(411,1424)(452,1424)
\thicklines \path(2276,1424)(2235,1424)
\put(370,1424){\makebox(0,0)[r]{ 0.4}}
\thicklines \path(411,1571)(452,1571)
\thicklines \path(2276,1571)(2235,1571)
\put(370,1571){\makebox(0,0)[r]{}}
\thicklines \path(411,1718)(452,1718)
\thicklines \path(2276,1718)(2235,1718)
\put(370,1718){\makebox(0,0)[r]{ 0.5}}
\thicklines \path(411,249)(411,290)
\thicklines \path(411,1718)(411,1677)
\put(411,166){\makebox(0,0){ 0}}
\thicklines \path(598,249)(598,290)
\thicklines \path(598,1718)(598,1677)
\put(598,166){\makebox(0,0){}}
\thicklines \path(784,249)(784,290)
\thicklines \path(784,1718)(784,1677)
\put(784,166){\makebox(0,0){ 0.1}}
\thicklines \path(971,249)(971,290)
\thicklines \path(971,1718)(971,1677)
\put(971,166){\makebox(0,0){ }}
\thicklines \path(1157,249)(1157,290)
\thicklines \path(1157,1718)(1157,1677)
\put(1157,166){\makebox(0,0){ 0.2}}
\thicklines \path(1344,249)(1344,290)
\thicklines \path(1344,1718)(1344,1677)
\put(1344,166){\makebox(0,0){ }}
\thicklines \path(1530,249)(1530,290)
\thicklines \path(1530,1718)(1530,1677)
\put(1530,166){\makebox(0,0){ 0.3}}
\thicklines \path(1717,249)(1717,290)
\thicklines \path(1717,1718)(1717,1677)
\put(1717,166){\makebox(0,0){ }}
\thicklines \path(1903,249)(1903,290)
\thicklines \path(1903,1718)(1903,1677)
\put(1903,166){\makebox(0,0){ 0.4}}
\thicklines \path(2090,249)(2090,290)
\thicklines \path(2090,1718)(2090,1677)
\put(2090,166){\makebox(0,0){ }}
\thicklines \path(2276,249)(2276,290)
\thicklines \path(2276,1718)(2276,1677)
\put(2276,166){\makebox(0,0){ 0.5}}
\thicklines \path(411,249)(2276,249)(2276,1718)(411,1718)(411,249)
\put(-82,983){\makebox(0,0)[l]{{\large $r_{C}$ }}}
\put(1343,42){\makebox(0,0){\large $r_{EW}$ }}
\put(1343,1442){\makebox(0,0){\large Allowed region  }}
\put(1903,1792){\makebox(0,0){$\delta^{EW} = \delta^{T} $ case  }}
\thinlines \path(411,748)(411,748)(448,748)(486,748)(523,748)(560,748)(598,748)(635,748)(672,719)(709,719)(747,719)(784,719)(821,719)(859,719)(896,719)(933,748)(970,748)(1008,748)(1045,748)(1082,748)(1120,778)(1157,778)(1194,778)(1232,807)(1269,807)(1306,807)(1344,807)(1381,837)(1418,837)(1455,837)(1493,837)(1530,837)(1567,837)(1605,837)(1642,866)(1679,866)(1716,866)(1754,866)(1791,895)(1828,895)(1866,895)(1903,895)(1940,925)(1978,925)(2015,925)(2052,954)(2090,954)(2127,954)(2164,984)(2201,984)(2239,1013)
\thinlines \path(2239,1013)(2276,1013)
\end{picture}
\eec
\end{minipage} 
\caption{The lower bound of $r_{C}$ to satisfy $R_c-R_n, S,$
    and $R_+ -2 $ at $1\sigma $ 
    at $r_T = 0.2 $ and $40^\circ < \phi_3 < 80^\circ $ 
    under constraint $-0.128 < A_{CP}^{+-} < -0.090$. The left figure
    shows that the case $\delta^{EW}$ and $\delta^C $ are free
    parameter and the right one is under constraint $\omega =
    \delta^{EW} - \delta^T =0$ and $\delta^C $ is still free
    parameter.  }
\label{REW-RC}
\end{center}
\end{figure}
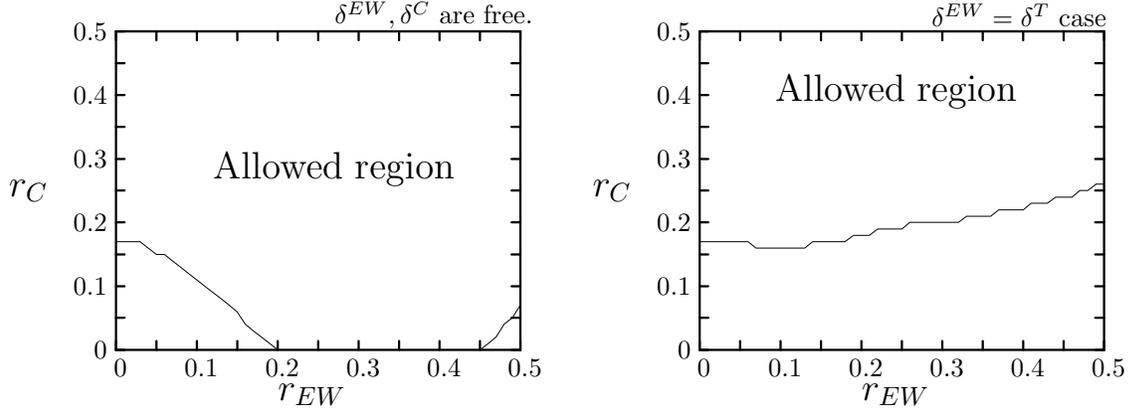

When we keep the terms with $r_C$ up to $O(r^2)$, 
To satisfy the relations about $R_c-R_n$, $S$ and $R_+ -2$, 
if $r_{EW}$ can not be so large, 
at least, $r_{C}$ should be large in spite of $r_{EW}$. 
Using the experimental bounds for $R_c-R_n$, $S$ and $R_+ -2 $, the
lower bound of $r_{C}$ for $r_{EW}$ are plotted in
Fig.\ref{REW-RC} in the cases $\delta^{EW}, \delta^C$ are free
parameters ( left ) and under constraint $\omega =\delta^{EW} - \delta^T = 0
$ (right). On both figures, the line show the lower bound to satisfy the
each relations at $1\sigma $ level. 
From the left figure, we find that in the small $r_{EW} $ case, 
larger $r_{C}$ is requested but it seems be too large because the
usual theoretical estimation is $r_{C} \simeq 0.02 $.  For
$r_{EW}=0.14$, $r_{C} $ should be larger than about $0.08$. 
If we put a constraint $\omega = \delta^{EW} - \delta^T = 0 $
as we discussed before, still larger $r_{C} $ will be favored.  
Thus there is also a possibility to explain by large $r_{C}$ 
but the magnitude might be still large compared with the usual
estimation. And it comes from tree diagram so that it may be slightly 
difficult to explain why $r_{C}$ is so large even if we consider some
new physics contribution. The 2 cases 
as the solutions by $r_{EW}$ or $r_C$ are summarized in Table 1.  
If so large $r_C$ 
which mean the magnitude is almost same with $r_T$ is allowed, 
it may help to explain the discrepancies for 
the branching ratios and direct CP asymmetries\cite{GROROS}.  
As a possibility, 
we discussed $r_C$ contribution after relaxing the hierarchy
assumption but we will discuss about new physics
contribution including in the penguin type diagrams. 
              
\begin{table}[htbp]
\begin{center}
\begin{tabular}{|c|c|c|c|c|}
\hline
parameter & theory 
            & $K\pi$ & $\pi\pi$ & $K\pi$ and $\pi\pi$ with $\omega =0$ \\
\hline
$r_{EW}$ & 0.14  & $ > 0.2$ & + New Phases  & $>0.2 $ \\
         &       &          & + SU(3) Breaking & + New Phases \\
         &       &          &                  & + SU(3) Breaking \\
\hline
$r_C$ & 0.02 & $> 0.08 $ & $ > 0.14 $  & $ > 0.18 $ \\
\hline
\end{tabular}
\caption{ Two solutions to solve the discrepancies in $B\rightarrow
K\pi$ and $B\rightarrow \pi \pi $. The solution by $r_C$ shows that 
the magnitude of $r_C$ has to be quite large. }
\label{table1}
\end{center}
\end{table}

If the deviations come from new physics contribution, it
has to be included in the penguin like contribution with new weak
phases because it is very difficult to produce such large strong phase 
difference as $\omega = \delta^{EW}-\delta^T \sim 100^\circ $ within the
SM. 
$B\rightarrow K\pi $ decays need large EW penguin contributions so
that it may be including the new physics contribution with new
weak phase in the EW penguin. 
Besides, the effect must appear also in the direct CP asymmetries. 
For example, $A_{CP}^{K^0\pi^0} \propto 2 r_{EW} \sin\delta^{EW}
\sin\theta^{New} $, so that we will have to check carefully these
modes. 

We consider a possibility of new physics in the penguin 
contributions. The ratios among the parameters with new phases 
are redefined as followes: $
\frac{T\, V_{ub}^*V_{us} }{P\, V_{tb}^* V_{ts} } \equiv 
     r_T\, e^{i\delta^T} e^{i (\phi_3+\theta^P) }\;,  
\frac{P_{EW} V_{tb}^*V_{ts} }{P\, V_{tb}^* V_{ts} } \equiv 
     r_{EW}\, e^{i\delta^{EW}} e^{i (\theta^{EW}+ \theta^P ) }\;, $ 
where $\theta^P$ and $\theta^{EW}$ are the weak phases coming from 
new physics. Using this parameterization, 
\bea
R_c - R_n  &=&
   2 r_{EW}^2 \left( 1 - 2 \cos^2\delta^{EW}
   \cos^2(\theta^{EW}+\theta^P) 
              \right)
    - 2 r_{EW}~r_T \cos(\delta^{EW}-\delta^{T})\cos(\phi_3 -
   \theta^{EW})
                \nn \\
& &  + 4 r_{EW} r_{T} 
       \cos\delta^{EW}\cos\delta^T \cos(\phi_3+\theta^P)
           \cos(\theta^{EW}+\theta^P ) .
\eea 
Because of the new weak phase $\theta^{EW}$ and $\theta^P$, the
constraints for the strong phases is fairly relaxed. 
The constraint for $r_{EW}$ is almost same but one for the strong
phases is changed and almost region for the strong phase
$\delta^{EW}$ will be allowed. Namely, small $\omega$ 
is also allowed in this case. In other words, the constraint 
for $\delta^{EW}$ is replaced to one for the new weak phase and their 
magnitude is not negligible value.    
To avoid the large strong phase difference, the EW penguin
must have new weak phase. 

When we respect the allowed region for 
the parameters in $B\rightarrow K \pi$, then 
they could not satisfy $B\rightarrow \pi \pi$ modes under the SU(3) 
symmetry. To explain the
both modes at once, SU(3) breaking effects in gluonic and EW penguin
diagrams with new phase will be strongly requested. 
In consequence, the role of the EW penguin contribution
will be more important even in $B\rightarrow \pi \pi$ modes. 
If there is any new physics and the effects appear 
through the loop effect in these modes, 
$B\rightarrow K \pi$ and $B\rightarrow \pi \pi$, will be helpful
modes to examine and search for the evidence of new physics. 
At the present situation, the deviation from the SM in $B\rightarrow K
\pi$ is still within the 
$2 \sigma$ level if large strong phase difference is allowed. Thus 
we need more accurate experimental data. 
In near future, we can use these modes to test the SM. 
For this purpose, the project the $B$ factories are upgraded\cite{LOI}
is helpful and important.

\medskip 

The work of S.M. was supported by the Grant-in-Aid for Scientific Research
in Priority Areas from the Ministry of Education, Culture, Science,
Sports and Technology of Japan (No.14046201). The work of T.Y. was  
supported by 21st Century COE Program of Nagoya University provided 
by JSPS.

\end{document}

%% file: fpcp04-macro.tex
\def\beq{\begin{equation}}
\def\eeq#1{\label{#1}\end{equation}}
\def\eeqn{\end{equation}}

\def\beqa{\begin{eqnarray}}
\def\eeqa#1{\label{#1}\end{eqnarray}}
\def\eeqan{\end{eqnarray}}

\let\bar=\overbar

\def\Dslash{\not{\hbox{\kern-4pt $D$}}}
\def\dslash{\not{\hbox{\kern-2pt $\del$}}}

\def\msb{{\bar{\ssstyle M \kern -1pt S}}}

\def\BB0bar{B^0 {\overline B}^0}
\def\BB0dbar{B_d^0 {\overline B}_d^0}
\def\BB0sbar{B_s^0 {\overline B}_s^0}

\RequirePackage{xspace}

\usepackage{relsize}
\def\babar{\mbox{\slshape B\kern-0.1em{\smaller A}\kern-0.1em
    B\kern-0.1em{\smaller A\kern-0.2em R}}}

\def\Kbar  {\kern 0.2em\overline{\kern -0.2em K}{}\xspace}

\def\Kz    {\ensuremath{K^0}\xspace}
\def\Kzb   {\ensuremath{\Kbar^0}\xspace}
\def\KzKzb {\ensuremath{\Kz \kern -0.16em \Kzb}\xspace}
\def\Kp    {\ensuremath{K^+}\xspace}
\def\Km    {\ensuremath{K^-}\xspace}

\def\KpKm  {\ensuremath{\Kp \kern -0.16em \Km}\xspace}

\def\Dbar    {\kern 0.2em\overline{\kern -0.2em D}{}\xspace}

\def\Dz      {\ensuremath{D^0}\xspace}
\def\Dzb     {\ensuremath{\Dbar^0}\xspace}
\def\DzDzb   {\ensuremath{\Dz {\kern -0.16em \Dzb}}\xspace}
\def\Dp      {\ensuremath{D^+}\xspace}
\def\Dm      {\ensuremath{D^-}\xspace}

\def\DpDm    {\ensuremath{\Dp {\kern -0.16em \Dm}}\xspace}

\def\Bbar    {\kern 0.18em\overline{\kern -0.18em B}{}\xspace}

\def\BB      {\ensuremath{B\Bbar}\xspace} 
\def\Bz      {\ensuremath{B^0}\xspace}
\def\Bzb     {\ensuremath{\Bbar^0}\xspace}
\def\BzBzb   {\ensuremath{\Bz {\kern -0.16em \Bzb}}\xspace}
\def\Bu      {\ensuremath{B^+}\xspace}
\def\Bub     {\ensuremath{B^-}\xspace}

\def\BpBm    {\ensuremath{\Bu {\kern -0.16em \Bub}}\xspace}

\mathchardef\Upsilon="7107
\def\Y#1S{\ensuremath{\Upsilon{(#1S)}}\xspace}% no space before {...}!

\mathchardef\Deltares="7101
\mathchardef\Xi="7104
\mathchardef\Lambda="7103
\mathchardef\Sigma="7106
\mathchardef\Omega="710A

\def\Deltabar{\kern 0.25em\overline{\kern -0.25em \Deltares}{}\xspace}
\def\Lbar{\kern 0.2em\overline{\kern -0.2em\Lambda\kern 0.05em}\kern-0.05em{}\xspace}
\def\Sigbar{\kern 0.2em\overline{\kern -0.2em \Sigma}{}\xspace}
\def\Xibar{\kern 0.2em\overline{\kern -0.2em \Xi}{}\xspace}
\def\Obar{\kern 0.2em\overline{\kern -0.2em \Omega}{}\xspace}
\def\Nbar{\kern 0.2em\overline{\kern -0.2em N}{}\xspace}
\def\Xb{\kern 0.2em\overline{\kern -0.2em X}{}\xspace}

\newcommand{\tev}{\ensuremath{\mathrm{\,Te\kern -0.1em V}}\xspace}
\newcommand{\gev}{\ensuremath{\mathrm{\,Ge\kern -0.1em V}}\xspace}
\newcommand{\mev}{\ensuremath{\mathrm{\,Me\kern -0.1em V}}\xspace}
\newcommand{\kev}{\ensuremath{\mathrm{\,ke\kern -0.1em V}}\xspace}
\newcommand{\ev}{\ensuremath{\mathrm{\,e\kern -0.1em V}}\xspace}
\newcommand{\gevc}{\ensuremath{{\mathrm{\,Ge\kern -0.1em V\!/}c}}\xspace}
\newcommand{\mevc}{\ensuremath{{\mathrm{\,Me\kern -0.1em V\!/}c}}\xspace}
\newcommand{\gevcc}{\ensuremath{{\mathrm{\,Ge\kern -0.1em V\!/}c^2}}\xspace}
\newcommand{\mevcc}{\ensuremath{{\mathrm{\,Me\kern -0.1em V\!/}c^2}}\xspace}
\def\mus  {\ensuremath{\rm \,\mus}\xspace}

\def\mus        {\ensuremath{\,\mu{\rm s}}\xspace}    %% microsecond
\def\pep2{PEP-II}

\def\gsim{{~\raise.15em\hbox{$>$}\kern-.85em
          \lower.35em\hbox{$\sim$}~}\xspace}
\def\lsim{{~\raise.15em\hbox{$<$}\kern-.85em
          \lower.35em\hbox{$\sim$}~}\xspace}

\xspace

\def\jetset74   {\mbox{\tt Jetset \hspace{-0.5em}7.\hspace{-0.2em}4}\xspace}